\documentclass{PoS}

\usepackage{cleveref}
\usepackage{gensymb}
\usepackage{graphicx}
\usepackage{subcaption}
\usepackage{multirow}
\usepackage{multicol}
\usepackage{url}

\newcommand{\plm}{$\pm$}

\newcommand{\veritas}{{VERITAS}}

\newcommand{\LAT}{\textit{Fermi}-LAT}

\newcommand{\GR}{$\gamma$-ray}

\title{Observing FRB 121102 with VERITAS; Searching for Associated TeV Emission}

\ShortTitle{Observing FRB 121102 with VERITAS}

\author{\speaker{Ralph Bird}, for the \veritas\ Collaboration\thanks{http://veritas.sao.arizona.edu}\\
        University of California, Los Angeles\\
        E-mail: \email{ralphbird@astro.ucla.edu}}

\abstract{Fast radio bursts are bright, unresolved and short flashes of radio emission originating from outside the Milky Way. The origin of these mysterious outbursts is unknown, but their high luminosity and short duration has prompted much speculation. The discovery that FRB 121102 repeats has enabled multiwavelength follow up, which has identified the host galaxy. VERITAS has observed the location of FRB 121102, including coincident observations with Arecibo. We present the results of a search for steady very high energy \GR\ emission and the methodology for searching for short timescale, transient optical and very high energy \GR\ emission.}

\FullConference{35th International Cosmic Ray Conference\\
		 10-20 July, 2017\\
		 Bexco, Busan, Korea}

\begin{document}

\section{Introduction}
Fast radio bursts (FRBs) are millisecond-duration burst of high dispersion-measure (DM) radio emission.
They were first discovered in 2007 in an analysis of archival data from 2001 July 24 taken by the Parkes observatory \cite{Lorimer}.
Subsequently many more bursts have been observed and these have enabled an understanding of their origin start to develop.

Their high DMs and uniform distribution across the sky implies that they originated outside the Milky Way and are thus orders of magnitude brighter than pulses from Galactic pulsars \cite{Thornton}.
Models were developed based upon arguments of causality which require that the emission region is sufficiently small to produce the observed bursts.
The fastest recorded bursts are less than 1~ms and not temporally resolved, leading to emission region sizes less than a few tens of kilometers across.
The properties of the bursts are also highly non-thermal, with brightness temperatures reaching up to about 10$^{37}$~K: in the region of temperatures measured from ``nanoshots'' from pulsars.
They were initially thought to be cataclysmic events; for example, the merger of neutron stars.
The discovery that FRB 121102 repeats ruled out all of these models (or at least rules out those models for that source. No other bursts have yet been seen to repeat, though FRB 110220 and FRB 140514 may form a pair \cite{Piro}).
The only astronomical phenomena with similar timescales that repeat are soft gamma repeaters (which have been observed to have rise times of 200-300~$\mu$s) and pulses, sub-pulses and ``nanoshots'' from radio pulsars, some of which have durations $\leq$~0.4~ns.
The properties and potential origin of FRBs is discussed in detail in \cite{Katz}.

\section{FRB 121102}
FRB 121102 is unique amongst FRBs in that it has been observed to repeat.
This surprising discovery was first reported in \cite{Spitler} and has lead to a large observing campaign to localize the origin and characterize the properties of the bursts.
Tendulkar et al. \cite{Tendulkar} identified the host as a dwarf Spheroidal galaxy located at (05$^h$31'58.70'' +33\degree 08'52.5'') and at a redshift of z=0.19273(8) (corresponding to a luminosity distance of 972~Mpc) where a persistent radio source has been detected.
The rate of bursts has been studied by Oppermann and Pen \cite{Oppermann}.
They reported that the repetition pattern is non-Poissonian; rather, they fit it with a Weibull distribution with a mean repetition rate of 5.7$^{+3.0}_{-2.0}$ per day and index, k, of 0.34$^{+0.06}_{-0.05}$ where a k of 1 reduces the distribution to a Poisson distribution and the observed value ($<$ 1) shows that smaller intervals between bursts are favored.
Multi-telescope campaigns in radio (e.g. \cite{Law}) and other wavelengths including \textit{Swift}, \textit{Chandra} and \textit{Fermi}-LAT \cite{Scholz} have identified over 100 bursts and been used to place upper limits on both the burst and steady state X-ray and high energy \GR\ fluxes.

\subsection{Very High Energy \GR\ Emission from FRB 121102}
As non-thermal emitters at very high luminosities (reaching 10$^{39}$~--~10$^{41}$~ergs, FRBs could potentially power very high energy (VHE, E $>$ 100~GeV) \GR\ emission.
At a distance of z=0.19273, VHE emission from FRB 121102 will be attenuated by interactions with the extragalactic background light (EBL).
Using the EBL model of \cite{Dominguez} we estimate an optical depth ($\tau$) of 0.02 at 100~GeV and 0.95 at 1~TeV.
Thus we do not expect to detect a significant flux above 1~TeV but emission in the region of 100~GeV will be largely unaffected by the EBL.
Models for VHE emission from FRBs exist.
Lyubarsky \cite{Lyubarsky} generated a model based upon soft gamma ray repeaters and predicted that millisecond VHE emission could be visible at distances up to about 100~Mpc.
In contrast, Vieyro et al. \cite{Vieryro} proposed a model for FRB 121102 based upon an AGN and concluded that high energy emission may be detectable for seconds to minutes after the radio burst, even for modest energy budgets.
However, the current theoretical and observational understanding of the emission of VHE \GR s from FRBs is very limited and VHE observations could be useful to constrain future modeling work by helping to constrain the total energy budget and e.

\subsection{Optical Emission from FRB 121102}
\label{sec:optBurst}
Designed to detect nanosecond pulses of Cherenkov radiation from relativistic particle showers in the Earth's atmosphere, imaging atmospheric Cherenkov telescopes (IACTs) are, by design, very efficient detectors of fast optical transients.
They have previously been used to conduct optical SETI searches \cite{VERITASoseti} and to study micro-meteorites \cite{WhMicroM}.
Lyutikov and Lorimer \cite{Lyutikov} predict that, assuming the same radio efficiency as optical efficiency, FRBs will produce millisecond flashes of magnitude, m, given by

\begin{equation}
m = -2.5 log_{10}\left( \frac{F}{3631 \mathrm{Jy}} \right)
\end{equation}
The brightest peak flux (F) from FRB 121102 reported in \cite{Law} is 3.3~Jy, which equates to a visual magnitude of 7.6.
They also comment that the radio emission is typically a small fraction of the total energy emitted at other wavelengths, it therefore not unreasonable to assume that the optical emission could be up to 100 times brighter than this, reaching a visual magnitude of about 2.6.

A search can be conducted using the same method as outlined in \cite{VERITASoseti}.
In short, the search technique looks for images which:
\begin{enumerate}
\item Appear at the source location in all four telescope cameras.
\item Have the same intensity in each telescope.
\item Are point-like.
\end{enumerate}

These criteria are sufficient to remove almost all of the background events from the analysis since cosmic ray showers produce the majority of their light below an altitude of 20 km.
Given that the telescopes are separated by about 100 m this means that the recorded images show significant parallactic displacement ($>$0.3\degree). 
In addition, cosmic (and $\gamma$) ray showers induce Cherenkov light over a significant vertical distance (km's).
Therefore, again due to the separation of the telescopes, all but (at most) one of the observed images will be elongated.

The temporal characteristics of the light pulse also play an important role in determining the trigger efficiency. 
The fastest bursts recorded thus far are $\lesssim 1$~ms in duration, though this is faster than the radio receivers temporal resolution so they may be shorter but, given that the majority of burst have widths of a few ms we would not expect them to have durations significantly less than this.
The \veritas\ PMT signal path is AC-coupled in the telescope camera prior to pre-amplification, with a lower cutoff frequency of approximately 100 kHz. 
This would prevent a pulse of the expected FRB duration from triggering the array and thus prevent \veritas\ from detecting such a pulse.

There is a caveat to this, in that a steady optical source such as a bright star in the field of view will generate high frequency Poisson noise fluctuations in the PMT signals, increasing the probability of accidentally crossing the discriminator thresholds. 
In \cite{VERITASoseti} they estimated that the number of photoelectrons produced by a B-band 6.4 mag star in the 12~ns readout window in a \veritas\ pixel is 18.8.
To trigger an individual pixel requires a pulse 4~--~5 photoelectrons \cite{Weinstein}.
With visual magnitudes potentially reaching up to 2.9, FRBs would produce around 140 photoelectrons within 3~ns, this is likely to be sufficient for the Poisson fluctuations to produce a trigger.

It is noted that an exact estimate of the sensitivity of the search would prove very difficult since the minimum detectable optical pulse intensity depends strongly upon the emission wavelength, and the duration and temporal profile of the pulse.
However, a detection would provide the first evidence of optical bursts associated with FRBs.

\section{\veritas\ Observations}
VERITAS is an array of four IACTs located at the Fred Lawrence Whipple Observatory (FLWO) in southern Arizona (31\degree\ 40'N, 110\degree\ 57'W, 1.3~km a.s.l.)\cite{VERITAS}.
Designed to detect the Cherenkov emission from extensive air showers produced by cosmic and \GR s, each telescope has a mirror area of 110~m$^2$ and is equipped with a 499-pixel camera of 3.5\degree\ diameter field of view with an angular resolution of 0.1\degree\ at 1~TeV. 
The system, completed in 2007, is run in a coincident mode requiring at least two of the four telescopes to trigger in each event. 
This design enables the observations of astrophysical objects in the energy range from 85~GeV to $>$~30~TeV with an effective area $>$~10$^5$~m$^2$. 
The observations were conducted wobble mode where the target is offset 0.5\degree\ from the center of the field-of-view to allow for simultaneous background estimation. 
The large effective area (over 10,000~m$^2$ at 1~TeV, cf. \LAT\ which reaches just less than 1~m$^2$ at 1~GeV \cite{LATPerf}) and low dead time ($<$~10\%, about 20~$\mu$s per event) combined with a low background rate (around 1 per minute within 0.1\degree\ of a given location after gamma/hadron selection cuts) makes \veritas\ an ideal candidate to search for fast, transient signals such as those that could be emitted from FRB 121102.

Initial observations were conducted in 2016 March, totaling 3.35 hours of live time, without contemporaneous radio coverage.
Additional observations were conducted in 2016 Fall and 2017 Winter which totaled 6.48 hours of live time and were conducted coincident with Arecibo observations.
In total, 10.83 hours of live time was taken on the source.

\section{Continuous VHE Emission}
The standard \veritas\ analysis techniques as described in \cite{LSI} were used to search for continuous TeV emission with the gamma/hadron selection cuts tuned to search for soft spectrum sources such as we would expect from FRB 121102 after EBL attenuation.

\begin{figure}[tb]
 \begin{center}
  \includegraphics[width=0.52\textwidth]{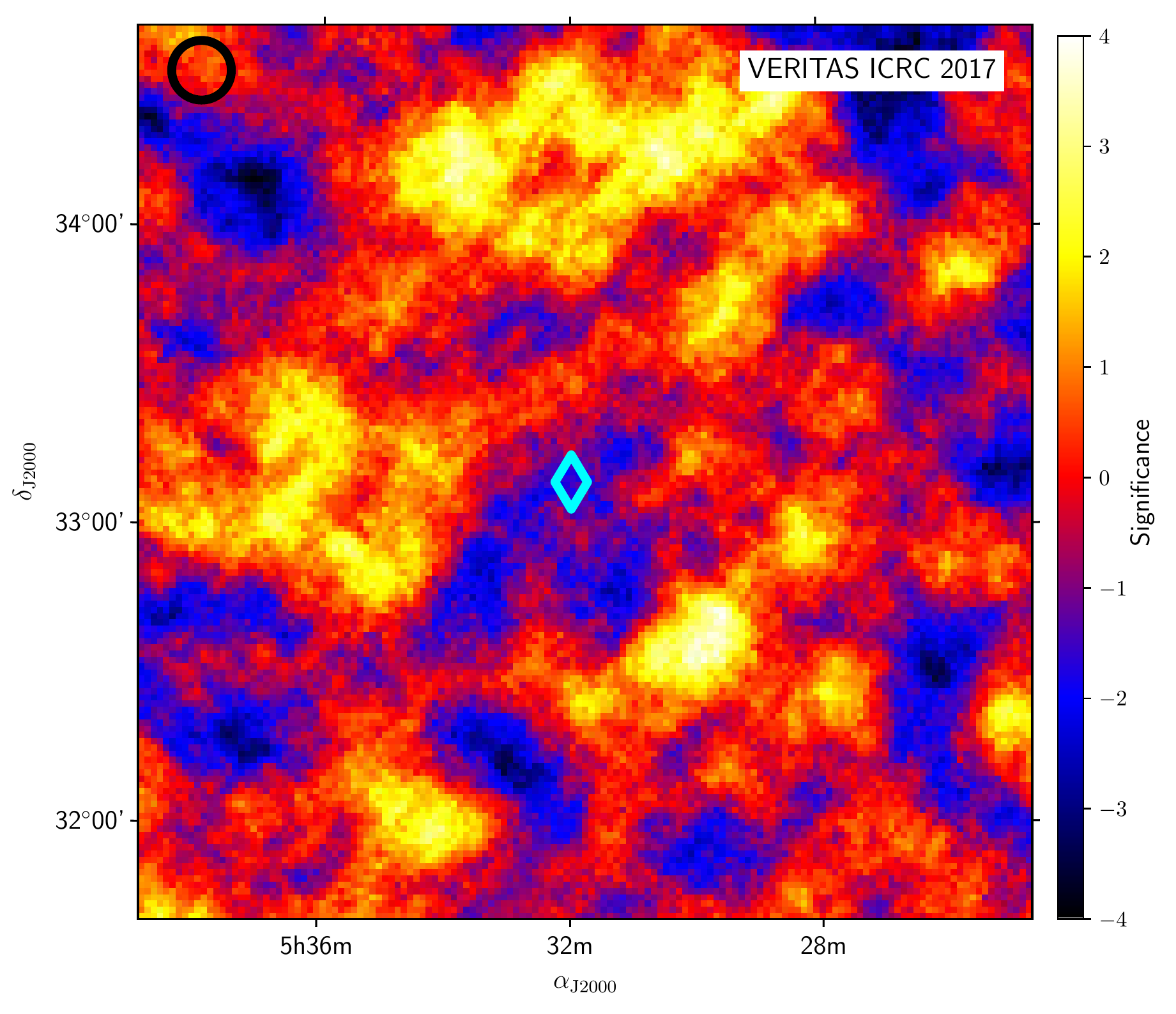}
  \hfill
  \includegraphics[width=0.465\textwidth]{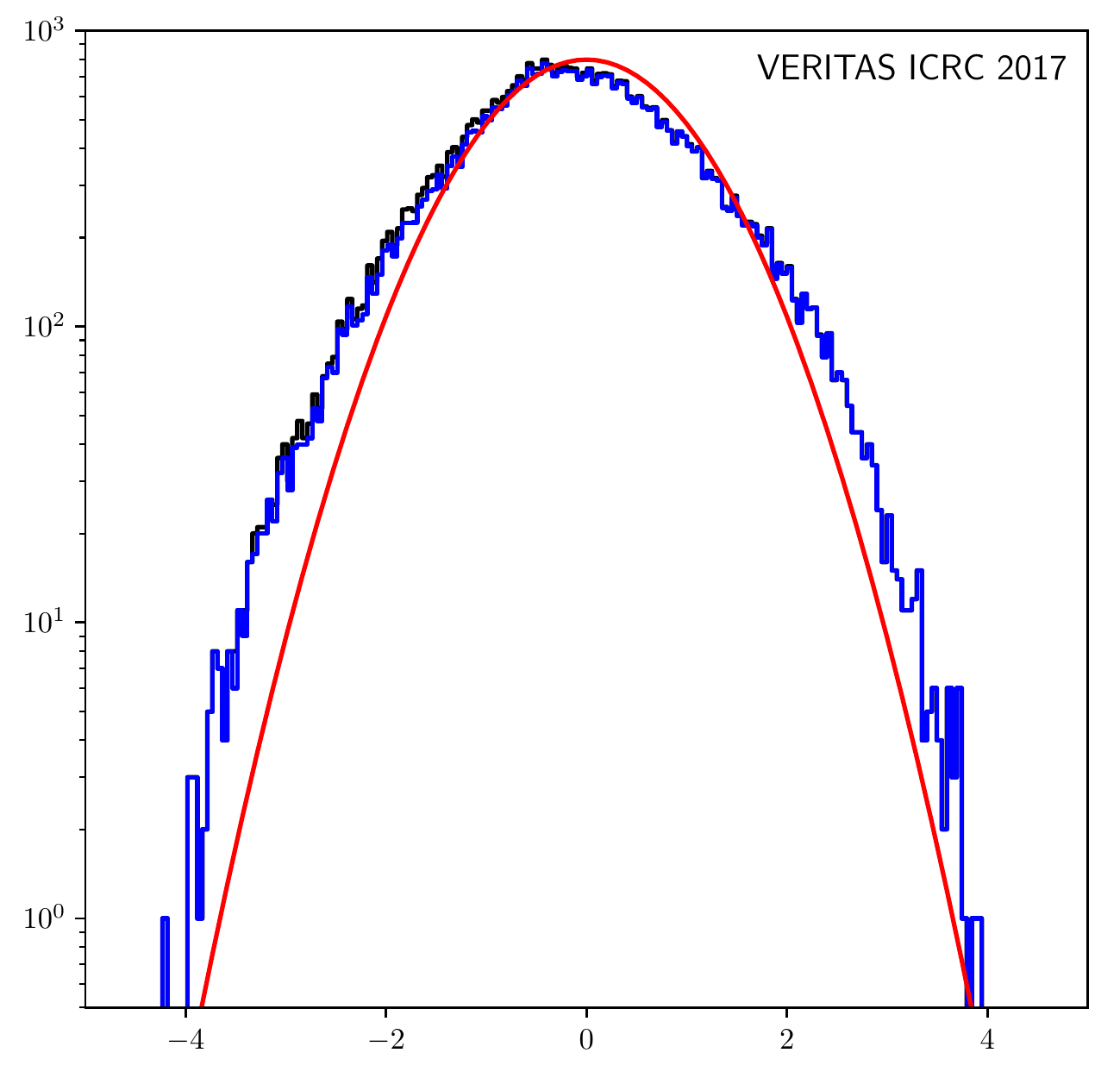}
 \end{center}
 \caption{ {\bf Left:}  Map of the significances for the region surrounding FRB 121102.  The location of FRB121102 is shown by the cyan diamond, the point spread function is shown by the black circle in the upper left corner
 {\bf Right:}  A histogram of the significances for the entire map.  Overlaid is a normal distribution (red line).
 \label{fig:SigMaps}}
\end{figure}

No VHE emission was detected from the position of FRB 121102 or the surrounding area.  
At the location of FRB 121102 the significance was -1.51$\sigma$.
A significance sky maps of the region is shown in \cref{fig:SigMaps}, along with a histogram of the significances, which is well fit by a normal distribution.
The mean significance was -0.086\plm 0.006 and the standard distribution was 1.106\plm 0.006.
This shows that no sources were detected in the field of view.

Given the non-detection of continuous VHE emission from FRB 121102, an upper limit on the flux was calculated using the method of Rolke \cite{Rolke} at the 95\% level and using the bounded solution.  
Differential flux upper limits were calculated using spectral indices of -2 and -4 at 5.2$\times 10^{-12}$ and 4.0$\times 10^{-11}$ cm$^{-2}$~s$^{-1}$~TeV$^{-1}$ at their respective energy thresholds of 0.2 and 0.15~TeV.
Here, the energy threshold is defined as the peak of the efficiency distribution: the effective area multiplied by a power law spectrum of -2 and -4 respectively.

Searches were also conducted on a run-by-run basis to search for emission on shorter time scales, no evidence of emission was found with a maximum significance in a single run of 2.1$\sigma$, before accounting for the statistical trials involved in searching many runs.
Differential upper limits were also calculated and are given in \cref{tab:runbyrun}.

\begin{table}
\centering
\caption{\veritas\ run by run results, showing, for each observing run, the start and end time, the live time (in minutes, after removal of periods of bad weather and accounting for deadtime), the significance of that runs observations, and, for two different spectral indices, the energy threshold (E$_{Thresh}$) in TeV and the differential upper limit at the energy threshold (in cm$^{-2}$~s$^{-1}$~TeV$^{-1}$).}
\label{tab:runbyrun}
\vspace{1mm}
\begin{tabular}{cccccccc}
\hline
\multirow{2}{*}{MJD Start} & \multirow{2}{*}{MJD End} & Live  & \multirow{2}{*}{Sig.} & \multicolumn{2}{c}{Index = -2} & \multicolumn{2}{c}{Index = -4} \\ 
&  &  Time & & E$_{Thresh}$ & DUL & E$_{Thresh}$ & DUL \\ \hline
57451.104884 & 57451.125718 & 23.78 & 0.31 & 0.14 & 1.11e-10 & 0.12 & 6.31e-10 \\
57451.1264 & 57451.147234 & 20.06 & -2.99 & 0.14 & 3.22e-11 & 0.13 & 1.58e-10 \\
57456.13441 & 57456.155243 & 25.89 & 0.57 & 0.15 & 8.96e-11 & 0.14 & 3.72e-10 \\
57456.156493 & 57456.177326 & 26.25 & 0.16 & 0.18 & 5.18e-11 & 0.17 & 2.26e-10 \\
57457.166262 & 57457.187106 & 25.89 & -0.25 & 0.18 & 4.26e-11 & 0.17 & 2.04e-10 \\
57457.18787 & 57457.208704 & 26.28 & -0.09 & 0.26 & 2.32e-11 & 0.24 & 9.07e-11 \\
57458.184028 & 57458.204861 & 26.39 & -1.17 & 0.24 & 1.58e-11 & 0.22 & 6.92e-11 \\
57459.17772 & 57459.198565 & 26.35 & 1.39 & 0.24 & 4.85e-11 & 0.22 & 2.05e-10 \\
57686.280347 & 57686.301192 & 14.08 & 0.1 & 0.38 & 8.77e-12 & 0.35 & 4.29e-11 \\
57686.301539 & 57686.322373 & 26.44 & 0.76 & 0.29 & 2.64e-11 & 0.26 & 1.21e-10 \\
57686.322882 & 57686.343715 & 25.93 & -0.31 & 0.24 & 2.27e-11 & 0.22 & 9.59e-11 \\
57686.344421 & 57686.365266 & 26.36 & -1.26 & 0.2 & 1.34e-11 & 0.18 & 5.34e-11 \\
57717.237118 & 57717.257951 & 27.05 & -0.5 & 0.22 & 3.2e-11 & 0.2 & 1.52e-10 \\
57717.258368 & 57717.279201 & 26.83 & 0.94 & 0.18 & 9.36e-11 & 0.17 & 4.84e-10 \\
57727.210486 & 57727.231319 & 25.61 & -1.48 & 0.22 & 1.9e-11 & 0.2 & 8.1e-11 \\
57727.231956 & 57727.252789 & 25.84 & -0.44 & 0.18 & 4.53e-11 & 0.17 & 1.89e-10 \\
57731.16684 & 57731.187685 & 25.79 & -0.26 & 0.32 & 9.35e-12 & 0.29 & 3.98e-11 \\
57731.188229 & 57731.209062 & 27.38 & 1.4 & 0.24 & 2.48e-11 & 0.2 & 1.44e-10 \\
57731.210127 & 57731.230961 & 27.53 & 0.46 & 0.18 & 2.75e-11 & 0.17 & 1.32e-10 \\
57731.231528 & 57731.241944 & 13.99 & 2.07 & 0.17 & 8.43e-11 & 0.15 & 3.78e-10 \\
57746.163796 & 57746.184641 & 26.94 & -0.7 & 0.22 & 3.15e-11 & 0.18 & 1.96e-10 \\
57746.185081 & 57746.19059 & 6.86 & 1.08 & 0.18 & 2.37e-10 & 0.17 & 1.24e-09 \\
57746.197743 & 57746.208218 & 10.26 & -0.91 & 0.15 & 8.16e-11 & 0.14 & 3.54e-10 \\
57772.085938 & 57772.106771 & 25.67 & -0.14 & 0.22 & 3.51e-11 & 0.2 & 1.54e-10 \\
57772.107315 & 57772.128148 & 25.72 & 0.0 & 0.18 & 5.52e-11 & 0.17 & 2.5e-10 \\ \hline
\end{tabular}
\end{table}

A $\sim$0.2~mJy persistent radio source \cite{Chatterjee} has been detected that is both compact ($<$ 0.7~pc) and coincident with FRB 121102 \cite{Marcote}.
At this distance, all of the known VHE \GR\ galactic sources would not be visible, being several orders of magnitude below the sensitivity of VERITAS.
Taking the Crab Nebula as a model (chosen due to its potential similarity to the host of the the FRB source) and scaling the VHE \GR\ luminosity by the ratio of its radio luminosity and the radio luminosity of the persistent counterpart to FRB 121102 (4$\times10^{5}$ \cite{Chatterjee}) would give a VHE luminosity of 7$\times 10^{35}$~s$^{-1}$~TeV$^{-1}$, or an observed flux of 3$\times 10^{-15}$cm$^{-2}$~s$^{-1}$~TeV$^{-1}$ at 0.2~TeV.
This is approximately three orders of magnitude lower than the current upper limit (ignoring any attenuation in the VHE \GR\ flux by the EBL).

\section{Pulsed VHE Emission}
Since FRBs, by their very nature, are not continuous emitters, an analysis looking for bursts of VHE emission is highly desirable.
This can be conducted in one of two ways:
\begin{description}
\item[Blind:] A search is conducted without any knowledge of whether a burst occurred.  This technique requires searching for an excess in emission within a given time window above the expected background rate.
The background rate is typically estimated from the data by scrambling either the positions or the times of the observed events and performing the same search, this can be repeated many times to reduce the statistical errors.
The proposed methodology is based upon the primordial black hole search conducted by Archambault \cite{Archambault} with the additional constraint that bursts must be originating from the direction of FRB 121102.
\item[Targeted:] Using knowledge of the occurrence of burst from simultaneous radio observations, a search is conducted around the time of a burst, looking for an excess in emission.  A similar technique is employed to estimate the background as for the blind analysis, however, since the time of the burst is known, the number of statistical trials is dramatically reduced. 
\end{description}
Given that bursts have not always been observed by all radio telescopes conducting concurrent observations, it is worth conducting a blind search in addition to the targeted search.  
However, to prevent bias, the targeted search is conducted first.
Since observations have been conducted both with and without simultaneous radio coverage a decision has been made to not conduct the blind search until the contemporaneous data has been analyzed.  
If bursts are detected in the targeted search then that knowledge can be used to reduce the number of different statistical tests which will be conducted (for example, the number different window sizes), increasing the statistical significance of any signal which is observed.
The results of the simultaneous Arecibo observations are not currently publicly available, thus the burst search has not yet been conducted.

\section{Pulsed Optical Emission}
As with the VHE burst search, an optical pulse search can be either conducted blind or targeted.
A blind search will look for any optical flashes that occur anywhere within the duration of the observations.
In contrast, a targeted search will look just in a small time window around any observed burst.
In order to not bias the results, as for the VHE burst search, the targeted optical burst search will be conducted prior to the blind burst search and after the Arecibo observations have been published.

\section{Conclusions}
10.83 hours (live time) of \veritas\ observations have been conducted on FRB 121102.
A steady state analysis of the region did not detect any evidence of VHE \GR\ emission and upper limits on the VHE flux of 5.2$\times 10^{-12}$ and 4.0$\times 10^{-11}$ cm$^{-2}$~s$^{-1}$~TeV$^{-1}$ at their energy thresholds  0.2 and 0.15~TeV were measured for spectral indices -2 and -4 respectively.
This is approximately three orders of magnitude above the VHE flux that would be expected assuming a Crab nebula like source scaled to match the luminosity of the persistent radio source located in the host galaxy.
Following the publication of the Arecibo observations we will conduct targeted and blind burst searches on the \veritas\ data, searching for both VHE and optical emission.

\section{Acknowledgements}
This research is supported by grants from the U.S. Department of Energy Office of Science, the U.S. National Science Foundation and the Smithsonian Institution, and by NSERC in Canada. We acknowledge the excellent work of the technical support staff at the Fred Lawrence Whipple Observatory and at the collaborating institutions in the construction and operation of the instrument. The VERITAS Collaboration is grateful to Trevor Weekes for his seminal contributions and leadership in the field of VHE gamma-ray astrophysics, which made this study possible.

\end{document}